\documentstyle[12pt]{article}

\advance \textheight by \topskip
\def\ps#1{\raisebox{.2ex}{$\displaystyle
    \mathop{\psi}^{\scriptscriptstyle [#1]}$}{}}
\def\xx#1{\raisebox{.2ex}{$\displaystyle
    \mathop{\xi}^{\scriptscriptstyle [#1]}$}{}} 
\def\eps#1{\raisebox{.2ex}{$\displaystyle
    \mathop{\varepsilon}^{\scriptscriptstyle (#1)}$}{}}
\def\H#1{\raisebox{.2ex}{$\displaystyle
    \mathop{H}^{\scriptscriptstyle [#1]}$}{}}
\def\Lam#1{\raisebox{.2ex}{$\displaystyle
    \mathop{\Lambda}^{\scriptscriptstyle [#1]}$}{}}
\def\Om#1{\raisebox{.2ex}{$\displaystyle
    \mathop{\Omega}^{\scriptscriptstyle [#1]}$}{}}
\def\Ph#1{\raisebox{.2ex}{$\displaystyle
    \mathop{\Phi}^{\scriptscriptstyle [#1]}$}{}}
\def\u#1{\raisebox{.2ex}{$\displaystyle
    \mathop{u}^{\scriptscriptstyle [#1]}$}{}}
\def\A#1#2{{\mathop{A}\limits^{[#1]}_{[#2]}}{}}
\def\th#1{\raisebox{.2ex}{$\displaystyle
    \mathop{\theta}^{\scriptscriptstyle [#1]}$}{}}
\def\c#1{\raisebox{.2ex}{$\displaystyle
    \mathop{c}^{\scriptscriptstyle (#1)}$}{}}
\def\pp#1{\raisebox{.2ex}{$\displaystyle
    \mathop{\pi}^{\scriptscriptstyle [#1]}$}{}}
\def\et#1{\raisebox{.2ex}{$\displaystyle
    \mathop{\eta}^{\scriptscriptstyle [#1]}$}{}}
\def\p#1{\raisebox{.2ex}{$\displaystyle
    \mathop{\bar{p}}^{\scriptscriptstyle [#1]}$}{}}
\def\F#1{\raisebox{.2ex}{$\displaystyle 
    \mathop{F}^{\scriptscriptstyle [#1]}$}{}}

\makeatletter
\@addtoreset{equation}{section}

\makeatother

\begin{document}

\begin{titlepage}
\hbox to \hsize{\hfil hep-th/9704183}
\hbox to \hsize{\hfil TUM--HEP--264/96}
\hbox to \hsize{\hfil December 1996}
\vfill
\large \bf
\begin{center}
The Ostrogradsky prescription for BFV formalism
\end{center}
\vskip1cm
\normalsize
\begin{center} 
Kh. S. Nirov{\footnote{E--mail: nirov@physik.tu-muenchen.de\hskip1cm
(Alexander von Humboldt fellow)}${}^{,}$
\footnote{On leave from the {\sl Institute for Nuclear Research of the 
Russian Academy of Sciences, Moscow}}}\\
{\small \it Institut f\"ur Theoretische Physik T30}\\
{\small \it Physik-Department, Technische Universit\"at M\"unchen}\\
{\small \it D-85747 Garching, Germany}
\end{center}
\vskip2cm
\begin{abstract}
\noindent 
Gauge-invariant systems of a general form with higher order
time derivatives of gauge parameters are investigated within
the framework of the BFV formalism. Higher order terms of the
BRST charge and BRST-invariant Hamiltonian are obtained.
It is shown that the identification rules for Lagrangian and
Hamiltonian BRST ghost variables depend on the choice of the
extension of constraints from the primary constraint surface.  
\end{abstract}
\vfill
\end{titlepage}

\section{Introduction}

For a consistent quantization of a system it is desirable to have 
its Hamiltonian description. If the system under consideration is
non-degenerate then this problem is in principle solved by the
Legendre transformation that is, at least locally, a bijective 
mapping of the velocity phase space onto the (canonical) phase space. 
In this, the quantization procedure endows the phase spaces with 
the sense of the state space of the system. 

But if the initial classical system is gauge-invariant then we have
no more one-to-one correspondence between the points of the phase 
spaces of the Lagrangian and the Hamiltonian descriptions of the system: 
actually, gauge invariance, in general, gives rise to constraints in 
these two formalisms because the Hessian is a degenerate matrix for the 
case, and so, the Legendre transformation turns out to be a singular 
mapping \cite{RS}. This circumstance forces us to modify the notion 
of the state space. The most powerful method to work out this situation 
in a covariant way is the BRST formalism based on the concept of the
BRST symmetry \cite{BRST}. 

There are two approaches to constructing a covariant BRST-invariant
effective theory. In the first one, called the Lagrangian BRST
formalism \cite{Lagr}, the initial object is a gauge-invariant
Lagrangian $L(q, \dot q)$ given on the corresponding velocity phase
space. Here, the gauge symmetry transformations
\[
\delta_\epsilon q^R = \epsilon^A \psi^R_A,
\]
where $\psi^R_A$ are some functionals of the trajectories of the
system, and $\epsilon^A$ are gauge parameters, are taken as a basis
to construct the so-called BRST transformations:
\[
\delta_\epsilon q^R \longrightarrow \delta_\lambda q^R =
\lambda s(q^R). 
\]
Here $\lambda$ is an odd parameter of (global) BRST transformations,
and $s$ is an odd vector field connected with the gauge
transformations in such a way that
\[
s(q^R) = c^A \psi_A^R, 
\]
and so that we have extended the initial configuration space of the 
system by adding globally defined ghost variables $c^A$ to initial
generalized coordinates $q^R$. Finally, we have to introduce the
so-called antighosts $\bar{c}_A$ that gives us the extended
configuration space and the corresponding velocity phase space of
the Lagrangian BRST formalism. 
In this, the ghost and antighost fields have the Grassmann parity 
being opposite to that the gauge parameters have:
\[
|c^A| = |\bar{c}_A| = |\epsilon^A| + \bar{1}. 
\]
Besides, while requiring that the initial generalized coordinates $q^R$
have the ghost number to be zero, we get that ghosts and antighosts
have ghost numbers which are opposite to each other:
\[
gh(q^R) = 0, \qquad gh(c^A) + gh(\bar{c}_A) = 0. 
\]
The effective BRST-invariant system is constructed on the extended
velocity phase space.

{}From the other side, we have the Hamiltonian BRST formalism elaborated
by Batalin, Fradkin and Vilkovisky \cite{BFV,Henn}. This one, called
also BFV formalism, begins with a Hamiltonian description of a system. 
In this we have a Hamiltonian $h$ and a set of irreducible constraints 
$\varphi_a$ which are functions being in involution.
Within the framework of the Hamiltonian BRST formalism we introduce
a new set of canonical pairs -- ghost variables $\eta^a$ and $\pi_a$
associated with the constraints $\varphi_a$, and put them to have
opposite ghost numbers:
\[
gh(\eta^a) + gh(\pi_a) = 0.
\]
The effective BRST-invariant theory is now defined on the extended 
phase space. The BRST transformations in the Hamiltonian approach 
are generated by the so-called BFV-BRST charge, that is a nilpotent 
odd operator globally defined on the extended phase space.

Thus, we arrive at a natural question: in which correspondence are
the ghost and antighost variables of the Lagrangian BRST formalism
and the ghost canonical pairs of the BFV formalism? This question
seems to be more intricate if we look at the following circumstance.
One can construct an equivalent Hamiltonian description of the
system when introducing another set of Hamiltonian and constraints,
i.~e. new functions $h'$ and ${\varphi_a}'$ which are some linear
combinations of $h$ and $\varphi_a$. This change of ingredients
of the Hamiltonian description of the system is equivalent, within
the framework of the BFV formalism, to a canonical transformation
of the extended phase space \cite{Henn}. And so, we get different
sets of the ghost canonical pairs, while in the Lagrangian BRST
formalism the ghost fields are fixed by the initial form of the
gauge transformations.  

Another question we consider in this paper, although being of a more 
technical character, is also connected with the above mentioned: 
this is the calculation of the higher order structure functions of 
the BFV formalism. In Refs.\cite{Sieg,MP} the authors investigated 
systems whose local symmetry transformations contain higher order 
time derivatives of gauge parameters. These systems are interesting 
from the physical point of view. Moreover, as is shown in Ref.\cite{RR}, 
the gauge algebra corresponding to the model of the so-called rigid 
particle \cite{MP} is equivalent to a particular case of $W$-algebras. 
Hence, it would be useful to have some general formulas providing BRST 
analysis of such systems "in advance". 

The paper is organized as follows. In Sec. 2 the Hamiltonian 
BRST-BFV formalism for gauge-invariant systems with higher order 
time derivatives of gauge parameters is constructed. 
Higher order structure functions of the corresponding BRST 
charge and BRST-invariant Hamiltonian are obtained. 
The Ostrogradsky prescription \cite{GT} is used to relate 
the BFV ghost canonical pairs with the ghost and antighost 
variables of the Lagrangian BRST formalism. In Sec. 3 it is 
shown that this relationship, called identification rules, 
depends on the choice of the extension of the constraints 
from the primary constraint surface.   
 
Summation over repeated indices is assumed.  All partial
derivatives are understood as the left ones \cite{DeW}.
Everywhere in the text integers within parentheses over
characters imply the corresponding order of time derivative,
while superscripts and subscripts within square brackets
just denote different functions.

\section{The BFV formalism and higher order structure functions}

Let a constrained system with a Hamiltonian $h$ and irreducible 
constraints $\varphi_a$ of the first class be given \cite{Dir}:
\begin{eqnarray}
\{\varphi_a,\varphi_b\} &=& f^c_{ab} \varphi_c, \label{2.1}\\
\{h,\varphi_a\} &=& h^b_a \varphi_b. \label{2.2}
\end{eqnarray}
We suppose the quantities entering Eqs.(\ref{2.1}),(\ref{2.2}) to be
globally defined on the phase space.

To provide the Hamiltonian description of the system within the framework
of the BFV formalism we have to construct the corresponding BFV-BRST charge
and the BFV-BRST-invariant Hamiltonian \cite{BFV,Henn}. These functions are
defined on the extended phase space and, in general, can be represented
as a series over the ghost variables. In this, the BRST charge is an odd
nilpotent operator having the ghost number $1$, and the BRST-invariant
Hamiltonian is defined as an even function with the ghost number $0$. 
 
For the constrained system of the first class introduced by Eqs.(\ref{2.1}) 
and (\ref{2.2}) we have the following general formulas. The BFV-BRST charge 
is given by the expression \cite{BFV,Henn}:
\begin{equation}
\Omega_B = \sum_{n \ge 0} \Om{n}_B = \sum_{n \ge 0}
\Om{n}^{b_1{\ldots}b_n}_{a_1{\ldots}a_{n+1}} \eta^{a_{n+1}}{\cdots}\,
\eta^{a_1} \pi_{b_n}{\cdots}\, \pi_{b_1} , \label{2.3}
\end{equation}
where
\begin{equation}
\Om{0}_{a_1} = \varphi_{a_1} , \label{2.4}
\end{equation}
and the quantities $\Om{n}^{b_1{\ldots}b_n}_{a_1{\ldots}a_{n+1}}$  
$(n > 0)$
are determined by the nilpotency condition \cite{BFV}
\begin{equation}
\{ \Omega_B\,,\,\Omega_B \} = 0 . \label{2.5}
\end{equation}
The BFV-BRST invariant Hamiltonian for the case can be written in the form
\begin{equation}
H_A = \sum_{n \ge 0} \H{n} = \sum_{n \ge 0}
\H{n}^{b_1{\ldots}b_n}_{a_1{\ldots}a_{n}} \eta^{a_n}{\cdots}\,
\eta^{a_1} \pi_{b_n}{\cdots}\, \pi_{b_1} . \label{2.6}
\end{equation}
Assuming that
\begin{equation}
\H{0} = h , \label{2.7}
\end{equation}
we can find the quantities $\H{n}^{b_1{\ldots}b_n}_{a_1{\ldots}a_n}$
from the BRST-invariance condition for $H_A$ \cite{BFV}
\begin{equation}
\{ H_A\,,\,\Omega_B \} = 0 . \label{2.8}
\end{equation}
The general theorem of the existence of the higher order structure 
functions
$\Om{n}^{b_1{\ldots}b_n}_{a_1{\ldots}a_{n+1}}$  and
$\H{n}^{b_1{\ldots}b_n}_{a_1{\ldots}a_n}$ of the BFV formalism is 
proved in Ref.\cite{Henn}. In particular, we have for $n = 1$
\begin{equation}
\Om{1}_B = - \frac{1}{2} f^c_{ab} \eta^b \eta^a \pi_c , \qquad
\H{1} = h^b_a \eta^a \pi_b . \label{2.9}
\end{equation}
For the second order structure functions ($n = 2$) we have the 
expressions \cite{BFV,Henn,NRjmp}:
\begin{eqnarray}
2 \Om{2}^{b_1c}_{a_1a_2a_3} \varphi_c &=& \frac{1}{6} \Bigl[
\{\varphi_{a_1},f^{b_1}_{a_2a_3}\} + \{\varphi_{a_2},f^{b_1}_{a_3a_1}\}
+ \{\varphi_{a_3},f^{b_1}_{a_1a_2}\} \nonumber \\
&& \;\; + f^c_{a_1a_2} f^{b_1}_{a_3c} + f^c_{a_2a_3} f^{b_1}_{a_1c}
+ f^c_{a_3a_1} f^{b_1}_{a_2c} \Bigr], \label{2.10} \\
&& \nonumber \\
2 \H{2}^{b_1c}_{a_1a_2} \varphi_c &=& \frac{1}{2} \Bigl[ 
\{h,f^{b_1}_{a_1a_2}\} - \{h^{b_1}_{a_1},\varphi_{a_2}\} +
\{h^{b_1}_{a_2},\varphi_{a_1}\} \nonumber \\
&& \;\; + h^c_{a_1} f^{b_1}_{a_2c} - h^c_{a_2} f^{b_1}_{a_1c} +
h^{b_1}_c f^c_{a_1a_2} \Bigr]. \label{2.11}
\end{eqnarray}
 
{}From the nilpotency of the BRST charge we get that Eq.(\ref{2.8})
defines $H_A$ only up to a BRST exact term. Hence, the general form
of the BRST invariant Hamiltonian is given by the expression
\begin{equation}
H_B = H_A - \{ \Psi\,,\,\Omega_B \} , \label{2.12}
\end{equation}
where $\Psi$ is an odd function, having the ghost number equal to 
$-1$. Thus, the gauge--fixing procedure within the framework of 
the BFV formalism consists in the choice of $\Psi$-function.

In Ref.\cite{N1} the Hamiltonian formalism was constructed for the 
gauge-invariant system given by the Lagrangian $L(q,\dot{q})$ with 
the gauge symmetry transformations of the form
\begin{equation}
\delta_\varepsilon q^r = \sum_{k=0}^N {\eps{k}^\alpha 
\ps{N-k}^r_\alpha (q,\dot q)},
\label{2.13}
\end{equation}
where $r = 1,\ldots,R$, $\alpha = 1,\ldots,A$ and $R > A$. 
In this we suppose the highest order $N$ of time derivatives 
of the gauge parameters to be more than $1$, and that the symmetry
transformations (\ref{2.13}) form a closed gauge algebra. 
Let us briefly recall some necessary definitions and formulas 
from Ref.\cite{N1}.  

Introduce $2R$--dimensional phase space with generalized coordinates 
$q^r$ and generalized momenta $p_r$, $r = 1,\ldots,R$, having the
Poisson bracket of the form:
\begin{equation}
\{ p_r\,,\,q^s \} = - \delta_r^s , \label{2.14}
\end{equation}
and define the mapping of the velocity phase space to this phase space
as usual: 
\begin{equation}
p_r(q,\dot q) = \frac{\partial L(q,\dot q)}{\partial \dot q^r} .
\label{2.15}
\end{equation}
{}From the gauge invariance of the system it follows that this mapping 
is singular. Actually, we have from the Noether identities that the
Hessian of the system has $A$ linearly independent null vectors
$\ps{0}^r_\alpha$. We suppose henceforth these vectors to be functions
of the generalized coordinates only. One can show that under the action
of the mapping (\ref{2.15}) one or several $A$--dimensional surfaces
having parametric representation of the form:
\begin{eqnarray}
q^r(\tau) &=& q^r, \label{2.16} \\
\dot{q}^r(\tau) &=& \dot{q}^r + \tau^\alpha \ps{0}^r_\alpha(q), 
\label{2.17}
\end{eqnarray}
are mapped into a point of the phase space. So, in our case, 
the image of the velocity phase space under the mapping (\ref{2.15}) 
is a $(2R - A)$--dimensional surface in the phase space, which may be 
defined by the following relations 
\begin{equation} 
\Ph{0}_\alpha (q,p) = 0 , \qquad \alpha = 1,\ldots,A , 
\label{2.18} 
\end{equation}
where the functions $\Ph{0}_\alpha$ are functionally independent.
By Eq.(\ref{2.18}) we have introduced the primary constraints of 
the system \cite{Dir} and, respectively, the primary constraint
surface \cite{RS}. Note that in the case under consideration, with
the null vectors of the Hessian being functions of the generalized
coordinates only, the primary constraints turn out to be linear in 
the generalized momenta \cite{NRjmp}. 

Let $F(q,p)$ be a function defined on the phase space. There
exists a function $f(q,\dot q)$ on the velocity phase space, 
such that
\begin{equation}
f(q,\dot q) = F(q,p(q,\dot q)) . 
\label{2.19}
\end{equation}
In this, the function  $f$ is constant on the surfaces given by 
Eqs.(\ref{2.16}),(\ref{2.17}). This fact is expressed by the 
differential equations of the form
\begin{equation}
\ps{0}^r_\alpha \frac{\partial f}{\partial \dot q^r} = 0 , 
\qquad \alpha = 1,\ldots,A . 
\label{2.20}
\end{equation}
We see however that not for any function $f(q,\dot q)$, given on the
velocity phase space, we can find a function $F(q,p)$ on the phase
space, which is connected with $f$ by the relation
\begin{equation}
F(q,p(q,\dot q)) = f(q,\dot q). \label{2.21}
\end{equation}
In this, Eq.(\ref{2.20}) gives the necessary condition for the 
existence of such a function $F(q,p)$. We assume these relations 
to be also sufficient for the validity of the corresponding 
Eq.(\ref{2.21}). It means that in our case any point of the 
primary constraint surface (\ref{2.18}) is the image of only 
one connected surface of the form (\ref{2.16}), (\ref{2.17}) 
\cite{RS}.

Therefore, for any function $f(q,\dot q)$ satisfying the equalities 
(\ref{2.20}) one can find a function $F(q,p)$, connected with $f$ 
by Eq.(\ref{2.21}) . We call such a function $f$ projectable to the 
primary constraint surface, or simply projectable, and write
\begin{equation}
F \doteq f . \label{2.22}
\end{equation}
We have by definition
\begin{equation}
\Ph{0}_\alpha \doteq 0 , \label{2.23}
\end{equation}
an so, any function $F$ of the form
\begin{equation}
F = F_0 + F^\alpha \Ph{0}_\alpha , \label{2.24}
\end{equation}
where $F_0$ satisfies Eq.(\ref{2.22}) and $F^\alpha$ are some 
arbitrary functions, satisfies the relation (\ref{2.22}) 
as well. Indeed, this equation determines the function $F$
only on the primary constraint surface and its solution is 
defined up to a linear combination of the primary constraints. 
Hence, the expression (\ref{2.24}) gives the general solution of 
Eq.(\ref{2.22}). It means that the relation(\ref{2.22}) specifies
the values of the function $F$ at the points of the primary
constraint surface only, and this function can be extended from
this surface (\ref{2.23}) to the total phase space arbitrarily. 
Note that according to Eq.(\ref{2.24}) various extensions will
differ from each other in linear combinations of the primary
constraints \cite{RS}. 
 
In this paper we use the notion of the standard extension
introduced earlier in Ref.\cite{PR} (on the definition and 
some useful properties of this method see also Ref.\cite{NRjmp}). 
Remember that a function $F(q,p)$ is called standard if it 
satisfies the relation:
\begin{equation}
\chi^\alpha_r \frac{\partial F}{\partial p_r} = 0 , \qquad
\alpha = 1,\ldots,A , \label{2.25}
\end{equation}
where the vectors $\chi^\alpha_r(q)$ are dual to the null vectors
of the Hessian, i.~e. the matrix
\begin{equation}
v^\alpha_\beta(q) = \ps{0}^r_\beta(q) \chi_r^\alpha(q)  \label{2.26}
\end{equation}
is nonsingular.

For the Poisson bracket of two standard functions $F$ and $G$ we have
the expression \cite{PR}
\begin{equation}
\{F\,,\,G\} = \{F\,,\,G\}^0 + \frac{\partial F}{\partial p_r}
(\frac{\partial \chi^\alpha_r}{\partial q^s} - 
\frac{\partial \chi^\alpha_s}{\partial q^r}) 
\frac{\partial G}{\partial p_s} \Ph{0}_\alpha.
\label{2.27}
\end{equation}  
For a notational conveniency we suppose that the vectors 
$\chi^\alpha_r$ can be chosen in such a way that 
\begin{equation}
\frac{\partial \chi^\alpha_r}{\partial q^s} 
- \frac{\partial \chi^\alpha_s}{\partial q^r} = 0. 
\label{2.28}
\end{equation}
It is clear that this assumption is not very restrictive. 
Moreover, from consequences of the gauge algebra 
\cite{N1} we see that for the systems with $N > 2$ the vectors 
$\chi^\alpha_r(q)$ can always be chosen in such a way that 
Eq.(\ref{2.28}) is fulfilled. 

Now the constraint algebra corresponding to the gauge-invariant 
system under consideration is given by the expressions \cite{N1}:  
\begin{eqnarray}
\{ \Ph{0}_\alpha\,,\,\Ph{0}_\beta \} &=& 0, 
\label{2.29} \\
&& \nonumber \\
\{ \Ph{k}_\alpha\,,\,\Ph{0}_\beta \} &=& \u{0}^\delta_\beta\,
\A{1}{N-k+1}^\gamma_{\alpha \delta} \; \Ph{1}_\gamma,
\label{2.30} \\
&& \nonumber \\
\{ \Ph{k}_\alpha\,,\,\Ph{l}_\beta \} &=& \Biggl( \u{k}^\delta_\alpha\,
\A{1}{N-l+1}^\gamma_{\beta \delta} \,-\, \u{l}^\delta_\beta\,
\A{1}{N-k+1}^\gamma_{\alpha \delta} \,+\, 
\dot q^r \frac{\partial}{\partial q^r}
\,\A{N-l}{2N-k-l}^\gamma_{\alpha \beta} \Biggr)^0\; 
\Ph{1}_\gamma \nonumber \\
&& \nonumber \\
&& +\, \sum_{i=0}^2 \sum_{j=0,1} {\left( \begin{array}{c}
                                     2N-k-l-i \\ N-l-j
                                    \end{array} \right)}
\A{j}{i}^\gamma_{\alpha \beta}\;\Ph{k+l-N+i}_\gamma,
\label{2.31} \\
&& \nonumber \\
\{ H\,,\,\Ph{0}_\alpha \} &=& \u{0}^\beta_\alpha \; \Ph{1}_\beta,  
\label{2.32} \\
&& \nonumber \\
\{ H\,,\,\Ph{k}_\alpha \} &=& \Ph{k+1}_\alpha -
\Biggl( \u{k}^\beta_\alpha \,+\, \mu^\delta 
\A{1}{N-k+1}^\beta_{\alpha \delta} \Biggr)^0\; \Ph{1}_\beta, 
\label{2.33}
\end{eqnarray}
where $k, l = 1,\ldots,N > 1$; $i > N - k - l$, and we use the 
notations 
\begin{equation}
v^\gamma_\alpha \u{0}_\gamma^\beta = \delta_\alpha^\beta, \qquad
\mu^\alpha = \dot q^r \chi_r^\beta \u{0}^\alpha_\beta, \qquad 
\u{k}^\alpha_\beta = \ps{k}^r_\beta
\chi_r^\gamma \u{0}^\alpha_\gamma.
\label{2.34}
\end{equation}
The symbol $f^0$ denotes the standard Hamiltonian counterpart of 
a function $f(q,\dot q)$ satisfying the relation (\ref{2.20}).
In this, the secondary constraints $\Ph{k}_\alpha$ and the 
Hamiltonian $H$ are the standard functions uniquely defined 
by the relations connecting them, respectively, with the 
Lagrangian constraints and the energy function of the system \cite{N1}:
\begin{equation}
\Ph{k}_\alpha \doteq \Lam{k}_\alpha, \qquad
H \doteq E = \dot{q}^r \frac{\partial L}{\partial \dot{q}^r} - L. 
\label{2.35}
\end{equation} 

Thus, we have the first class constraint system and can apply
to it the general BFV formalism \cite{BFV,Henn} in order to
construct a BRST-invariant effective theory. To this end, let 
us enlarge the initial phase space by adding to the canonical 
pairs of even variables $q^r$, $p_r$ the set of odd ghost
coordinates $\et{k}^\alpha$, associated with the constraints of the
system, and ghost momenta $\pp{k}_\alpha$, $k = 0,\ldots,N$, 
$\alpha = 1,\ldots,A$, endowed with the ghost numbers, respectively, 
$1$ and $-1$. We suppose the nonzero Poisson brackets for the ghost 
variables to be of the form
\begin{equation}
\{ \pp{k}_\alpha\,,\,\et{l}^\beta \} = -\,
\delta^{kl} \delta^\beta_\alpha , \qquad k, l = 0,\ldots,N . 
\label{2.36}
\end{equation}

The BFV-BRST charge $\Omega_B$ and the BFV-BRST-invariant Hamiltonian $H_A$ 
are constructed on the extended phase space according to the general
BFV prescriptions Eqs.(\ref{2.3})-(\ref{2.8}). Using the constraint
algebra in the standard extension we obtain for $\Omega_B$:
\begin{eqnarray}
\Om{0}_B &=& \sum_{k=0}^N \et{k}^\alpha \Ph{k}_\alpha, 
\label{2.37}\\
&& \nonumber \\
\Om{1}_B &=& \u{0}^\delta_\alpha 
\left( \A{1}{1}^\gamma_{\beta \delta}\,\et{N}^\beta
\,+\, \A{1}{2}^\gamma_{\beta \delta}\,\et{N-1}^\beta \right) 
\et{0}^\alpha\,\pp{1}_\gamma \nonumber\\
&& \nonumber \\ 
&& -\, \frac{1}{2} \sum_{k,l=1}^N \left( \u{k}^\delta_\alpha\,
\A{1}{N-l+1}^\gamma_{\beta \delta} \,-\, \u{l}^\delta_\beta\,
\A{1}{N-k+1}^\gamma_{\alpha \delta} \,+\,
\dot q^r \frac{\partial}{\partial q^r}\,
\A{N-l}{2N-k-l}^\gamma_{\alpha \beta} \right)^0\;
\et{l}^\beta\,\et{k}^\alpha\,\pp{1}_\gamma \nonumber \\
&& \nonumber \\
&& -\, \frac{1}{2} \sum_{k,l=1}^N \sum_{i=0}^2 \sum_{j=0,1}
                                 {\left( \begin{array}{c}
                                         2N-k-l-i \\ N-l-j
                                         \end{array} \right)}
\A{j}{i}^\gamma_{\alpha \beta}\;\et{l}^\beta\, \et{k}^\alpha\;
\pp{k+l-N+i}_\gamma, \label{2.38} 
\end{eqnarray}
and for $H_A$:
\begin{eqnarray}
\H{0} &=& H, \label{2.39}\\
&& \nonumber\\ 
\H{1} &=& \et{0}^\alpha \, \u{0}^\beta_\alpha \, \pp{1}_\beta 
\,+\, \sum_{k=1}^N \et{k}^\alpha \left[ \pp{k+1}_\alpha \,-\,
\left( \u{k}^\beta_\alpha \,+\, \mu^\delta
\A{1}{N-k+1}^\beta_{\alpha \delta} \right)^0\,\pp{1}_\beta \right].
\label{2.40}
\end{eqnarray}

{}From Eqs.(\ref{2.10}) and (\ref{2.11}) we see that to get the next order 
terms of $\Omega_B$ and $H_A$ we need the expressions for the Poisson 
brackets of the constraints and the Hamiltonian with the structure 
functions of the constraint algebra. 
The necessary formulas were obtained in Ref.\cite{N1}. 
Recall that the corresponding calculations are based
on the notion of the pseudoinverse matrix \cite{Lan-R}. 
In this we define the projector
\begin{equation}
\Pi^r_s = \delta^r_s - \chi^\alpha_s \u{0}^\beta_\alpha \ps{0}^r_\beta,
\qquad \Pi^t_s \Pi_t^r = \Pi^r_s, \label{2.41}
\end{equation}
and then the pseudoinverse matrix $W^{rs}$, corresponding to the Hessian
of the system, is uniquely defined by the relations
\begin{equation}
W^{rt} W_{ts} = \Pi^r_s, \qquad W^{rs} \chi_s^\alpha = 0. 
\label{2.42}
\end{equation}
So, for an arbitrary standard function $F(q,p)$ connected 
with a function $f(q,\dot q)$ by the relation (\ref{2.22}) 
we have for our case \cite{N1}:
\begin{eqnarray}
\{\Ph{0}_\alpha,F\} &\doteq& \u{0}^\beta_\alpha \, \xx{0}_\beta(f), 
\label{2.43}\\
\{\Ph{k}_\alpha,F\} &\doteq& \xx{k}_\alpha(f) - \u{k}^\beta_\alpha\,
\xx{0}_\beta(f) + \Biggl( \frac{\partial f}{\partial \dot{q}^r} 
W^{rs} \frac{\partial \u{k}^\beta_\alpha}{\partial \dot{q}^s} \Biggr)\,
\Lam{1}_\beta, 
\label{2.44}\\
\{H,F\} &\doteq& - T(f) + \mu^\alpha \xx{0}_\alpha, 
\label{2.45}
\end{eqnarray}
where the vector fields $\xx{k}_\alpha$ are of the form
\begin{equation}
\xx{k}_\alpha = \ps{k}^r_\alpha \frac{\partial}{\partial q^r} + 
\left(\ps{k+1}^r_\alpha + T\Bigl(\ps{k}^r_\alpha\Bigr)\right)
\frac{\partial}{\partial \dot{q}^r}, \qquad
k = 0,1,\ldots,N,
\label{2.46}
\end{equation}
and we have used the notation
\begin{equation}
T = \dot{q}^t \frac{\partial}{\partial q^t} + R_s W^{st}
\frac{\partial}{\partial \dot{q}^t}. 
\label{2.47}
\end{equation}
Note that the differential operator $T$ has the sense of the evolution
operator of gauge-invariant systems \cite{Isp1-Isp2}:
\begin{equation}
T(f) \vert_{L_r = 0} = \frac{d}{dt}(f), \label{2.48} 
\end{equation}
and the algebra of the vector fields $\xx{k}_\alpha$, $k=0,1,\ldots,N$, 
and $\ps{0}_\alpha^r \frac{\partial}{\partial \dot{q}^r}$  
coincides with the gauge algebra \cite{N1} on the trajectories of 
the system.

Making use of Eqs.(\ref{2.43}) and (\ref{2.44}) and consequences of the
Jacobi identities \cite{N2} for the gauge transformations (\ref{2.13}) 
we obtain from Eq.(\ref{2.10}) that the only nonzero second order 
structure functions of the BRST charge are given by the expression:
\begin{eqnarray}
\Om{2}^{\sigma\tau}_{\mu\nu\rho}(k,l,m) &=& \frac{1}{12} \Biggl( 
\frac{\partial \u{k}^\tau_\mu}{\partial\dot{q}^r} W^{rs}
\frac{\partial(\ps{l}^p_\nu \Pi^t_p)}{\partial\dot{q}^s}
\frac{\partial \u{m}^\sigma_\rho}{\partial\dot{q}^t} + 
\frac{\partial \u{m}^\tau_\rho}{\partial\dot{q}^r} W^{rs}
\frac{\partial(\ps{k}^p_\mu \Pi^t_p)}{\partial\dot{q}^s}
\frac{\partial \u{l}^\sigma_\nu}{\partial\dot{q}^t} \nonumber \\
&& \nonumber\\
&& \hspace{0.7cm} + 
\frac{\partial \u{l}^\tau_\nu}{\partial\dot{q}^r} W^{rs}
\frac{\partial(\ps{m}^p_\rho \Pi^t_p)}{\partial\dot{q}^s}
\frac{\partial \u{k}^\sigma_\mu}{\partial\dot{q}^t}
\Biggr)^0 , \quad k,l,m = 1,\ldots,N.
\label{2.49}
\end{eqnarray}
and so, we have that
\begin{equation}
\Om{2}_B = \frac{1}{4} \sum_{k,l,m=1}^N \Biggl(
\frac{\partial \u{k}^\tau_\mu}{\partial\dot{q}^r} W^{rs}
\frac{\partial(\ps{l}^p_\nu \Pi^t_p)}{\partial\dot{q}^s}
\frac{\partial \u{m}^\sigma_\rho}{\partial\dot{q}^t} \Biggr)^0
\;\et{m}^\rho\,\et{l}^\nu\,\et{k}^\mu\,\pp{1}_\sigma\,\pp{1}_\tau.
\label{2.50}
\end{equation}
Further, using additionally Eq.(\ref{2.45}) we get from Eq.(\ref{2.11})
that the only nonzero structure functions of the second order of the
BRST-invariant Hamiltonian are 
\begin{equation}
\H{2}^{\sigma\tau}_{\mu\nu}(k,l) = - \frac{1}{4} \Biggl(
\frac{\partial \u{k}^\sigma_\mu}{\partial\dot{q}^r} W^{rs}
\frac{\partial \u{l}^\tau_\nu}{\partial\dot{q}^s} - 
\frac{\partial \u{l}^\sigma_\nu}{\partial\dot{q}^r} W^{rs}
\frac{\partial \u{k}^\tau_\mu}{\partial\dot{q}^s}
\Biggr)^0, \qquad k,l = 1,\ldots,N. 
\label{2.51}
\end{equation}
{}From the last relation we get immediately
\begin{equation}
\H{2} = - \frac{1}{2} \sum_{k,l=1}^N \Biggl(
\frac{\partial \u{k}^\tau_\mu}{\partial\dot{q}^r} W^{rs}
\frac{\partial \u{l}^\sigma_\nu}{\partial\dot{q}^s} \Biggr)^0\;
\et{l}^\nu\,\et{k}^\mu\,\pp{1}_\sigma\,\pp{1}_\tau.
\label{2.52} 
\end{equation}

The expressions for the structure functions of the BFV formalism 
can be obtained also in the Lagrangian approach making use of the 
techniques elaborated in Ref.\cite{NRjmp}.
This way is more convenient when the initial object is a 
gauge-invariant Lagrangian. For the system under consideration
one needs first to apply the Ostrogradsky prescription to the
corresponding BRST-invariant Lagrangian $L_B$ \cite{N2} 
\begin{equation}
L_B = L \,-\, \frac{1}{2} \chi^\alpha \gamma_{\alpha \beta} \chi^\beta 
\,-\,
\bar{c}_\alpha s(q^r) \chi^\alpha_{;r} \,+\, \dot{\bar{c}}_\alpha s(q^r)
\frac{\partial \chi^\alpha}{\partial \dot q^r} ,
\label{2.53}
\end{equation}
where the notation $\chi^\alpha_{;r}$ for the variation of the functions 
$\chi^\alpha(q,\dot{q}) = \dot{q}^r \chi_r^\alpha(q) + \nu^\alpha(q)$ 
over the trajectory $q^r(t)$ is used. One of the results we get 
in this way is the identification rules for the BFV ghost canonical
pairs and ghosts and antighosts introduced in the Lagrangian formalism. 
Namely, we introduce new odd variables putting 
\begin{equation}
\th{k}^\alpha = \c{N-k}^\alpha, \qquad k = 1,\ldots,N. \label{2.54}
\end{equation}
Now according to the Ostrogradsky formalism the extended configuration 
space is described by the set of even and odd variables 
$q^r$, $\bar{c}_\alpha$ and $\th{k}^\alpha$. 
Using the Ostrogradsky prescription we define the generalized momenta,
corresponding to the system with the Lagrangian $L_B$, by the formulas
\cite{N2}:
\begin{eqnarray}
p_r &=& \frac{\partial L_B}{\partial \dot{q}^r} = 
\frac{\partial L}{\partial \dot{q}^r} - 
\chi_r^\alpha \gamma_{\alpha\beta} \chi^\beta - \bar{c}_\alpha
\frac{\partial s(q^t)}{\partial \dot{q}^r} \chi^\alpha_{;t} 
+ \dot{\bar{c}}_\alpha \frac{\partial s(q^t)}{\partial \dot{q}^r} 
\chi_t^\alpha,
\label{2.55} \\
&& \nonumber \\
p^\alpha &=& \frac{\partial L_B}{\partial \dot{\bar{c}}_\alpha} =
\sum_{k=0}^N \c{k}^\beta \ps{N-k}^r_\beta \chi_r^\alpha, 
\label{2.56}\\
&& \nonumber \\
\p{k}_\alpha &=& \sum_{l=1}^k (-1)^{k-l} \frac{d^{k-l}}{dt^{k-l}}
\left( \frac{\partial L_B}{\partial \c{N-l+1}^\alpha}\right) =
\sum_{l=1}^k (-1)^{k-l} \frac{d^{k-l}}{dt^{k-l}}
\left[ \ps{l-1}^r_\alpha 
\left( \chi^\beta \bar{c}_\beta \right)_{;r} \right],
\nonumber \\
&& \qquad {\rm for}\;\;{\rm any}\;\;k = 1,\ldots,N. 
\label{2.57}
\end{eqnarray}
The Hamiltonian description of the system with the Lagrangian $L_B$ is
constructed on the phase space with canonically conjugate variables
$q^r$, $\bar{c}_\alpha$, $\th{k}^\alpha$ and $p_r$, $p^\alpha$, 
$\p{k}_\alpha$. 
When comparing the forms of the BRST charge obtained from the 
Lagrangian and the Hamiltonian BRST formalisms, we obtain the 
following correspondence between the odd variables:
\begin{equation}
\et{0}^\alpha = p^\alpha, \qquad 
\et{k}^\alpha = \th{k}^\alpha, \qquad 
\pp{0}_\alpha = \bar{c}_\alpha, \qquad 
\pp{k}_\alpha = \p{k}_\alpha.
\label{2.58}
\end{equation}
However it is important to note that the BFV ghosts $\et{k}^\alpha$ 
are associated with the standard constraints and the relations 
(\ref{2.58}) should be modified for different extensions from the 
primary constraint surface. We deal with this question in the next 
section.

\section{On the canonical transformations}

In Ref.\cite{Henn} it was shown that a transition from a set
of constraints to another one can be realized in the Hamiltonian
BRST formalism, at least locally, as a canonical transformation 
of the extended phase space. In order to demonstrate how this
statement works in our case, let us consider a simplified situation
with the vectors $\ps{k}^r_\alpha$, $k=0,1,\ldots,N$, being 
functions of the generalized coordinates only. 

One can easily find that the gauge algebra in this case is
equivalent to the following Lie algebra of the vector fields
$\ps{k}_\alpha = \ps{k}^r_\alpha \frac{\partial}{\partial q^r}$:
\begin{equation}
\bigl[\,\ps{k}_\alpha\,,\,\ps{l}_\beta\,\bigr] = 
{\left( \begin{array}{c} 
          2N-k-l \\ N-l
        \end{array} \right)} 
\A{0}{0}^\gamma_{\alpha\beta} \ps{k+l-N}_\gamma, \label{3.1}
\end{equation}
where $k,l=0,1,\ldots,N$ and the only nonzero structure functions
$\A{0}{0}^\gamma_{\alpha\beta}$ are constant. 

{}From the corresponding constraint algebra we immediately get the 
expression of the BRST charge:
\begin{equation}
\Omega_B = p^\alpha \Ph{0}_\alpha 
+ \sum_{k=1}^N \th{k}^\alpha \Ph{k}_\alpha
+ \frac{1}{2} \sum_{k,l=1}^N {\left( \begin{array}{c}
                                        2N-k-l \\ N-l
                                     \end{array} \right)}
\A{0}{0}^\gamma_{\alpha\beta}\,
\th{k}^\alpha\,\th{l}^\beta\,\p{k+l-N}_\gamma. \label{3.2} 
\end{equation}
Remember that this form of the BRST charge (\ref{3.2}) corresponds 
to the standardly extended constraints $\Ph{k}_\alpha$, and we have
the identification rules (\ref{2.58}).

Let us now choose another set of the constraints. To this end, we
take into account that for any non-degenerate matrix $w^\beta_\alpha$
the equations
\begin{equation}
w^\beta_\alpha \, \Ph{0}_\beta = 0 \label{3.3}
\end{equation}  
define one and the same surface in the phase space -- the primary
constraint surface given by Eq.(\ref{2.18}), and consider the functions
\begin{equation}
\F{0}_\alpha = v^\beta_\alpha \, \Ph{0}_\beta \label{3.4}
\end{equation}
as new primary constraints. Besides, let us choose the following
non-standard extension for all other constraints: 
\begin{equation}
\F{k}_\alpha = \Ph{k}_\alpha 
+ \u{k}^\gamma_\alpha v^\beta_\gamma \Ph{0}_\beta. \label{3.5} 
\end{equation}
In Eq.(\ref{3.5}) we used that various extensions from the primary
constraint surface to the total phase space differ from each other
in a linear combination of the primary constraints. 

The BRST charge corresponding to the new constraint algebra is given
by the expression
\begin{equation}
\Omega_B = p^\alpha \F{0}_\alpha + \sum_{k=1}^N \th{k}^\alpha \F{k}_\alpha
+ \A{0}{0}^\varepsilon_{\alpha\beta} \u{0}^\gamma_\varepsilon\,
\th{N}^\alpha\,p^\beta\,\bar{c}_\gamma + \frac{1}{2} \sum_{k,l=1}^N
                   {\left( \begin{array}{c}
                             2N-k-l \\ N-l
                           \end{array} \right)}
\A{0}{0}^\gamma_{\alpha\beta}\,
\th{k}^\alpha\,\th{l}^\beta\,\p{k+l-N}_\gamma.
\label{3.6}
\end{equation}
While comparing the expressions (\ref{3.2}) and (\ref{3.6}) for the
BRST charge, it is easy to show that the transition from the set of
constraints $\Ph{0}_\alpha$, $\Ph{k}_\alpha$ to $\F{0}_\alpha$, 
$\F{k}_\alpha$ can be realized as the following canonical transformation
of the extended phase space:
\begin{eqnarray}
\bar{c}_\alpha &\rightarrow& v^\beta_\alpha \bar{c}_\beta, 
\label{3.7}\\
p^\alpha &\rightarrow& p^\beta \u{0}^\alpha_\beta + 
\sum_{k=1}^N \th{k}^\beta \u{k}^\alpha_\beta, 
\label{3.8}\\
\p{k}_\alpha &\rightarrow& \p{k}_\alpha - 
\u{k}^\gamma_\alpha v^\beta_\gamma \bar{c}_\beta,
\label{3.9}\\
p_r &\rightarrow& p_r - 
p^\alpha \frac{\partial\u{0}^\gamma_\alpha}{\partial q^r}
v^\beta_\gamma \bar{c}_\beta - \sum_{k=1}^N \th{k}^\alpha
\frac{\partial\u{k}^\gamma_\alpha}{\partial q^r} v^\beta_\gamma
\bar{c}_\beta. 
\label{3.10} 
\end{eqnarray} 

The canonical transformation (\ref{3.7})--(\ref{3.10}) obviously 
changes the identification rules (\ref{2.58}). To have the same 
correspondence between the new ghost variables, one needs to perform
an equivalent transformation in the Lagrangian formalism.
We see that the corresponding transformation consists of two steps:
first we have to redefine the antighosts in $L_B$ Eq.(\ref{2.53}) 
as follows:
\begin{equation}
\bar{c}_\alpha = \u{0}^\beta_\alpha \bar{c}^\prime_\beta, 
\label{3.11}
\end{equation} 
and then to construct a new BRST-invariant Lagrangian 
\begin{equation}
L_B \rightarrow L_B'' = L_B' - \frac{d}{dt} \sigma_{can},
\label{3.12}
\end{equation}
where we have used the notation
\begin{equation}
\sigma_{can} = \sum_{k=1}^N \th{k}^\alpha \u{k}^\beta_\alpha 
\bar{c}^\prime_\beta. \label{3.13}
\end{equation}
We see that redefinition of the primary constraints Eq.(\ref{3.4})
is related to redefinition of the antighosts Eq.(\ref{3.11}) 
and there is no generator of the corresponding global canonical 
transformation (\ref{3.7}) and certain parts of Eqs.(\ref{3.8}),
(\ref{3.10}). This is connected with the very property of the 
primary constraints, whose appearance is of a non-dynamical 
character. On the other hand, to the change of the standard
extension, obtained by the relation (\ref{3.5}), corresponds 
globally defined generator of the canonical transformation 
$\sigma_{can}$.  

One can also choose a non-standard extension for the 
Hamiltonian $H$ as follows:
\begin{equation}
H \rightarrow H + D^\alpha \Ph{0}_\alpha, \label{3.14}
\end{equation}
where $D^\alpha$ are some functions on the phase space. 
We easily find that the transition (\ref{3.14}) is equivalent
in the BFV formalism not to a canonical transformation, but to
the shift of the gauge--fixing fermion:
\begin{equation}
\Psi \rightarrow \Psi + \bar{c}_\alpha D^\alpha. \label{3.15}
\end{equation}

\section{Conclusion}

In this paper we have analyzed gauge-invariant systems with $N$-th
order time derivatives of gauge parameters within the framework 
of the Hamiltonian BRST-BFV formalism. Higher order structure functions
of the corresponding BFV-BRST charge and BFV-BRST-invariant Hamiltonian have 
been calculated on the basis of the results obtained earlier in 
Refs.\cite{NRjmp,N1,N2}. It has been shown that higher order terms 
for the systems with $N > 1$ are formally of the same form as for 
the systems with $N = 1$ \cite{NRjmp}: one needs only to take into 
account an additional summation over the ghosts associated with the
secondary constraints of $k$-th stage, $k = 1,\ldots,N$. 
The difference between these systems appears in the forms of the 
corresponding first order structure functions. Besides, for the 
systems with $N > 1$ we have an additional strong restriction on 
the structure functions of the (closed) gauge algebra \cite{N1}: 
for the case of $N = 2$ these structure functions depend on the 
generalized coordinates only, and for the systems with $N > 2$ 
they turn out to be constant, whereas for the case of $N = 1$ the 
structure functions of the gauge algebra may depend on all the 
velocity phase space coordinates. In this respect, the systems 
with higher order time derivatives of gauge parameters do not 
generalize the systems with $N = 1$, but present different 
classes of gauge-invariant systems.

Another principal property of the systems considered here is the
problem of identifying Lagrangian and Hamiltonian ghost variables.
This problem is solved with the use of the Ostrogradsky prescription. 
The corresponding identification rules turned out to be connected
with the choice of the extension of the constraints from the primary
constraint surface. 

It would be interesting to generalize the results of our consideration
to the systems, whose gauge symmetry transformations form an open gauge
algebra.

\vskip0.7cm
\small

{\bf Acknowledgements.}
\vskip0.3cm
The author is grateful to Prof. H. P. Nilles for stimulating discussions,
support and kind hospitality at his University. This work has been supported 
by the Alexander von Humboldt Fellowship and by the European Commission TMR 
programme ERBFMRX--CT96--0045 and ERBFMRX--CT96--0090.


\begin{thebibliography}{**}


\bibitem{RS}
A. V. Razumov and L. D. Soloviev, "Introduction to classical mechanics
of constrained systems," IHEP preprints 86-212, 86-213, 86-214
(Protvino, 1986, in Russian).

\bibitem{BRST}
C. Becchi, A. Rouet and R. Stora, {\it Phys. Lett.} {\bf B52}, 
344 (1974);
{\it Commun. Math. Phys.} {\bf 42}, 127 (1975); {\it Ann. Phys.} 
{\bf 98}, 287 (1976); 
I. V. Tyutin, "Gauge invariance in field theory and statistical
physics in operatorial formalism," FIAN preprint {\bf 39}
(Moscow, 1975, in Russian).

\bibitem{Lagr}
T. Kugo and S. Uehara, {\it Nucl. Phys.} {\bf B197}, 378 (1982);
F. R. Ore and P. van Nieuwenhuizen, {\it Nucl. Phys.} {\bf B204}, 317 
(1982); 
L. Alvarez--Gaum\'e and L. Baulieu, {\it Nucl. Phys.} {\bf B212}, 255 
(1983); 
L. Baulieu, {\it Phys. Rep.} {\bf 129}, 1 (1985).

\bibitem{BFV}
E. S. Fradkin and G. A. Vilkovisky, {\it Phys. Lett.} {\bf B55}, 
224 (1975); 
I. A. Batalin and G. A. Vilkovisky, {\it Phys. Lett.} {\bf B69}, 
309 (1977); 
E. S. Fradkin and T. E. Fradkina, {\it Phys. Lett.} {\bf B72}, 
33 (1978); 
I. A. Batalin and E. S. Fradkin, {\it Phys. Lett.} {\bf B122}, 
17 (1983); 
I. A. Batalin and E. S. Fradkin, {\it Riv. Nuovo Cimento} {\bf 9(10)}, 
1 (1986).

\bibitem{Henn}
M. Henneaux, {\it Phys. Rep.} {\bf 126}, 1 (1985).

\bibitem{Sieg}
W. Siegel, {\it Int. J. Mod. Phy.} {\bf A3}, 2713 (1988); 
I. Bakas and E. Kiritsis, {\it Phys. Lett.} {\bf B301}, 49 (1993); 
J. Ellis, N. E. Mavromatos and D. V. Nanopoulos, "Some physical
aspects of Liouville string dynamics,"  CERN-TH-7269-94 (hep-th/9405196).

\bibitem{MP}
M. S. Plyushchay, {\it Mod. Phys. Lett.} {\bf A4}, 837 (1989). 

\bibitem{RR}
E. Ramos and J. Roca, {\it Nucl. Phys.} {\bf B436}, 529 (1995);
{\it Nucl. Phys.} {\bf B452}, 705 (1995).  

\bibitem{GT}
D. M. Gitman and I. V. Tyutin, {\sc Canonical quantization of fields with
constraints}  (Springer-Verlag, Berlin Heidelberg, 1990). 

\bibitem{DeW}
B. DeWitt, {\it Supermanifolds} (Cambridge University Press,
Cambridge, Second Edition, 1992).  

\bibitem{Dir}
P. A. M. Dirac, {\sc Lectures on quantum mechanics}
(Yeshiva University, New York, 1964). 

\bibitem{NRjmp}
Kh. S. Nirov and A. V. Razumov, {\it J. Math. Phys.} {\bf 34}, 
3933 (1993).

\bibitem{N1}
Kh. S. Nirov, {\it Int. J. Mod. Phys.} {\bf A10}, 4087 (1995).

\bibitem{PR}
P. N. Pyatov and A. V. Razumov, {\it Int. J. Mod. Phys.} {\bf A4}, 
3211 (1989).

\bibitem{Lan-R}
P. Lancaster, {\sc Theory of matrices} (Academic, New York, 1969); 
A. V. Razumov, "Dependent coordinates in classical mechanics," 
IHEP preprint 84-86 (Protvino, 1984).

\bibitem{Isp1-Isp2}
C. Battle, J. Paris, J. M. Pons and N. Rom\'an-Roy, 
{\it J. Math. Phys.} {\bf 27}, 2953 (1986);
{\it Lett. Math. Phys.} {\bf 13}, 17 (1987);
X. Gr\'acia and J. M. Pons, {\it Lett. Math. Phys.} {\bf 17}, 175 (1989).  

\bibitem{N2}
Kh. S. Nirov, {\it Int. J. Mod. Phys.} {\bf A11}, 5279 (1996).

\end{thebibliography}
\end{document}